\documentclass[conference]{IEEEtran}
\IEEEoverridecommandlockouts

\usepackage{cite}
\usepackage{amsmath,amssymb,amsfonts}
\usepackage{algorithmic}
\usepackage{graphicx}
\usepackage{textcomp}
\usepackage{xcolor}
\usepackage{multirow}
\usepackage{subfigure}
\usepackage[colorlinks,
            linkcolor=blue,
            anchorcolor=blue,
            citecolor=blue,
            urlcolor=blue]{hyperref}
\usepackage{cleveref}
\newcommand{\figref}[2]{\hyperref[#2]{\ref*{#1}(\subref*{#2})}} 

\def\BibTeX{{\rm B\kern-.05em{\sc i\kern-.025em b}\kern-.08em
    T\kern-.1667em\lower.7ex\hbox{E}\kern-.125emX}}
\begin{document}

\title{\huge A Power-Efficient Hardware Implementation of \textit{L-Mul}}

\author{\IEEEauthorblockN{{Ruiqi Chen}, {Yangxintong Lyu}, {Han Bao}, {Bruno da Silva} \\Department of Electronics and Informatics (ETRO), Vrije Universiteit Brussel, Brussels, Belgium \\ Email:{\tt\small\{ruiqi.chen, yangxintong.lyu, han.bao, bruno.da.silva\}@vub.be} }} 

\maketitle

\begin{abstract}
Multiplication is a core operation in modern neural network (NN) computations, contributing significantly to energy consumption. The linear-complexity multiplication (\textit{L-Mul}) algorithm is specifically proposed as an approximate multiplication method for emerging NN models, such as large language model (LLM), to reduce the energy consumption and computational complexity of multiplications. However, hardware implementation designs for \textit{L-Mul} have not yet been reported. Additionally, 8-bit floating-point (FP8), as an emerging data format, offers a better dynamic range compared to traditional 8-bit integer (INT8), making it increasingly popular and widely adopted in NN computations.
This paper thus presents a power-efficient FPGA-based hardware implementation (approximate FP8 multiplier) for \textit{L-Mul}. The core computation is implemented using the dynamic reconfigurable lookup tables and carry chains primitives available in AMD Xilinx UltraScale/UltraScale+ technology. The accuracy and resource utilization of the approximate multiplier are evaluated and analyzed. Furthermore, the FP8 approximate multiplier is deployed in the inference phase of representative NN models to validate its effectiveness. 
\end{abstract}


\section{Introduction}

In recent years, the rapid development of neural network (NN) has brought the issue of high energy consumption to the forefront~\cite{dampfhoffer2023backpropagation}. An effective approach to addressing this challenge is the use of approximate computing and quantization techniques. Approximate computing simplifies key operations in neural networks, such as multiplication and other nonlinear functions. The emerging approximate multiply method, linear-complexity multiplication (\textit{L-Mul}) algorithm~\cite{luo2024addition}, has been deployed on GPUs for large language models (LLM), providing initial validation of its effectiveness. As for quantization techniques, they improve memory access and computational efficiency~\cite{han2020extremely, yao2022zeroquant}. Among these, 8-bit floating-point (FP8) numbers, compared to traditional 8-bit integer (INT8), offer better dynamic range and computational precision, making them widely adopted in neural network computations~\cite{shen2024efficient, lutz2024fused}. To enhance the computational efficiency of FP8, specialized hardware modules for FP8 acceleration have become a new trend. For instance, designing application-specific integrated circuits (ASICs)~\cite{lee20217, venkataramanaiah202328} and integrating corresponding units into GPUs~\cite{elster2022nvidia}. 

As a fundamental computation in DNNs, various approximate multipliers have been proposed to improve efficiency and reduce energy consumption.  These multipliers are designed to reduce latency, energy consumption, and area. Chen et al. ~\cite{chen2020optimally} proposed an optimally multi-level architecture that seamlessly integrates runtime configurability with parallel module execution. An optimization strategy was applied to improve area efficiency, achieving a linear relationship with accuracy rather than the quadratic or exponential relationships seen in previous works. Ansari et al.~\cite{ansari2020improved} developed an 8$\times$8 approximate multiplier tailored for NN designs by improving the design of logarithmic multipliers. HEAM~\cite{zheng2022heam} achieves automated design of approximate multipliers by minimizing the average error based on operand distribution and integrates these multipliers into DNN accelerators. However, the aforementioned approximation methods are primarily aimed at reducing power consumption and area utilization in ASIC implementations and may perform suboptimally on FPGAs. This is because FPGA reconfigurable logic is typically based on fixed-size lookup tables (LUTs). While FPGAs also integrate DSP hardware multiplier units, these units are physically fixed and limited in quantity. Therefore, improving the efficiency of LUT-based multiplication in terms of speed, power consumption, and resource utilization becomes particularly critical. Ullah et al.~\cite{ullah2018smapproxlib, ullah2020area, ullah2021high} proposed a series of FPGA-based approximate multipliers covering data bit-widths from 4-bit to 32-bit. More recently, their AxO series~\cite{liu2024bitsys, ullah2024axospike} integrated the design of approximate multipliers into SNN accelerators. DyRecMul~\cite{vakili2024dyrecmul} introduced a dynamically reconfigurable INT8 approximate multiplier design, which includes a floating-point conversion unit. This design enables efficient floating-point conversion, reducing preprocessing operations and enhancing computational efficiency. Leon et al.~\cite{leon2021improving} proposed a DSP-based approximate multiplier design for floating-point computations, which was integrated into a CNN accelerator. This approach achieved more efficient computation within the accelerator framework. 

Although previous works have made efforts in FPGA-based approximate multiplier designs, there is still a lack of specialized approximate multipliers targeting the FP8 format. Therefore, this paper present a power-efficient hardware implementation for \textit{L-Mul} and the main contributions include:
\begin{itemize}
    \item We implement \textit{L-Mul}, an approximate multiplication method for FP8 computations, on FPGA. Using LUT and carry-chain primitives, we achieve fine-grained optimization to minimize resource usage and power consumption. To the best of our knowledge, this is the first FPGA-based FP8 approximate multiplier design.
    \item We validate our design on AMD UltraScale/UltraScale+ devices. Compared to previous FPGA-based 8-bit approximate multiplier designs, our approach reduces resource consumption by an average of 10\% while maintaining comparable performance. Additionally, we integrate our design into a CNN accelerator, and experiments demonstrate that, among 8-bit designs, ours achieves the highest accuracy, energy efficiency, and the lowest latency.
\end{itemize}

\section{Preliminaries}
\subsection{FP8 Formats}
\label{sec_bg_fp8}
\begin{figure}[hp!] 
\centering   \includegraphics[width=\linewidth]{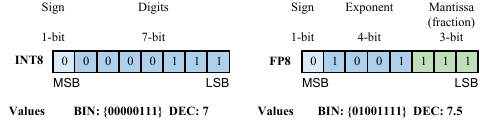}  
\caption{The demonstration of the INT8 and FP8 (E4M3) defined in IEEE 754. MSB stands for most significant bit and LSB stands for least significant bit.}   \label{fig_int8fp8}   
\end{figure}

FP8 is a natural progression from the FP16 representations, effectively reducing memory consumption and improving memory access and computational efficiency~\cite{lutz2024fused}. Compared to traditional INT8, FP8 offers a larger dynamic range (the commonly used E4M3 format, as shown in Table~\ref{tab_int8fp8} and Fig.~\ref{fig_int8fp8}). Moreover, FP8 delivers good results for neural network inference processes~\cite{van2023fp8}. The FP8 format adheres to IEEE-754 conventions, where a real numbers is encoded using a 1-bit sign $S$, an $e$-bit integer exponent $E$, and an $m$-bit fractional (mantissa $M$),
\begin{equation}
\label{eq_fp8}
x_{\text{DEC}}=(-1)^{S}\times2^{E}\times M,
\end{equation}
where $E=e-bias$ and $M=1+m$. The $bias$ in this context varies with the number of bits in the exponent, and it is determined by the following formula:
\begin{equation}
\label{eq_bias}
bias=2^{e-1}-1.
\end{equation}
Note that an implict 1, the $hidden bit$, is concatenated to the fraction as an integer bit, forming the significand. A FP number with $E$ = 0 has no implicit 1 in the significand so zero and subnormal values can be represented. In addition, exponent $E$=$2^{e-1}$$-1$ is reserved for the representation of $\pm\infty$ and $NaNs$.

\begin{table}[hp!]
\centering
\caption{Comparison of INT8 and FP8 (E4M3)}
\label{tab_int8fp8}
\begin{tabular}{|c|c|c|}
\hline
\textbf{Data Type}    & INT8 & FP8 (E4M3) \\ \hline
\textbf{Bit Width}    & 8 bits & 8 bits\\ \hline
\textbf{Minimum Value} & -128  & -448 \\ \hline
\textbf{Maximum Value} & 127 &  448 \\ \hline
\textbf{Decimal Precision}    & Fixed (1) & Dynamic \\ \hline
\end{tabular}
\end{table}

\subsection{FPGA Structure}
\label{sec_FPGA}
State-of-the-art FPGAs from AMD (Xilinx) and Altera (Intel) utilize basic logic cells such as 6-input LUTs, carry chains, multiplexers, and D flip-flops to implement both combinational and sequential logic circuits. The typical structure of AMD's FPGA is presented as an example for design implementation. However, the proposed methodology is generic and can be implemented on FPGAs from other vendors, including Altera and Anlogic, which also use 6-input LUTs, carry chains, multiplexers, and D flip-flops.

\begin{figure}[tp!]
    \centering
    \setlength{\abovecaptionskip}{-12pt}
    \renewcommand{\thesubfigure}{}
    \includegraphics[width=\linewidth]{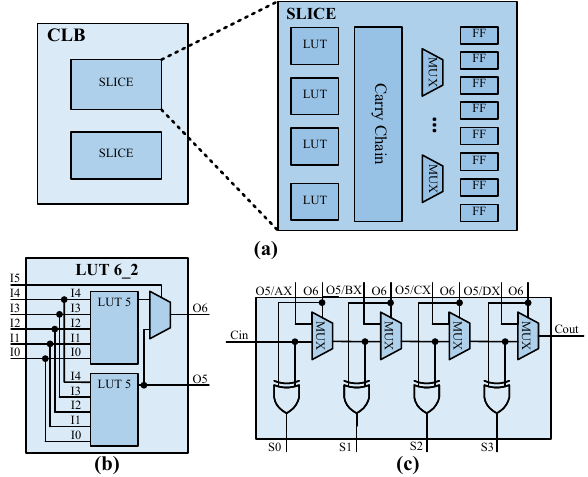}
    \subfigure[]{\label{fig_clb}}
    \subfigure[]{\label{fig_lut}}
    \subfigure[]{\label{fig_carry}}
    \caption{Typical AMD FPGA configurable logic block (CLB) structure~\cite{ultra}. 
    \hyperref[fig_clb]{(a)} Logic cell within the CLB. 
    \hyperref[fig_lut]{(b)} LUT6 structure. 
    \hyperref[fig_carry]{(c)} Carry chain structure.}
    \label{fig:main_structure}
\end{figure}

A slice in the configurable logic block (CLB) of AMD’s 7-series and UltraScale/UltraScale+ FPGAs contains four 6-input LUTs (commonly referred to as LUT6$\_$2), two 4-bit carry chains (a 8-bit carry chain within UltraScale/UltraScale+), eight flip-flops and multiplexers, as shown in Fig.~\ref{fig_clb}. A LUT6$\_$2 can be used to implement either a single 6-bit combinational function, using the O6 output bit, or two 5-bit combinational functions, using the O5 and O6 output bits, as shown in Fig.~\ref{fig_lut}. This is done by defining an INT value, which describes all the possible input combinations for which a logic value "1" is required at the output. For example, an INT value of 0000000000000002 (hex) for LUT6$\_$2 defines to produce outputs O5 = 1 and O6 = 0 for input combination 100001. Besides the implementation of single 6-bit combinational functions, these LUT6$\_$2 are also used for controlling the associated carry chain, as shown in Fig.~\ref{fig_carry}. The carry chain implements a carry-lookahead adder, using O5 as the carry-generate signal and O6 as the carry-propagate signal. The carry-generate signals for the carry chain can also be provided by the external bypass signals AX – DX.

\begin{figure*}[!tp]
\centering 
\includegraphics[width=\textwidth]{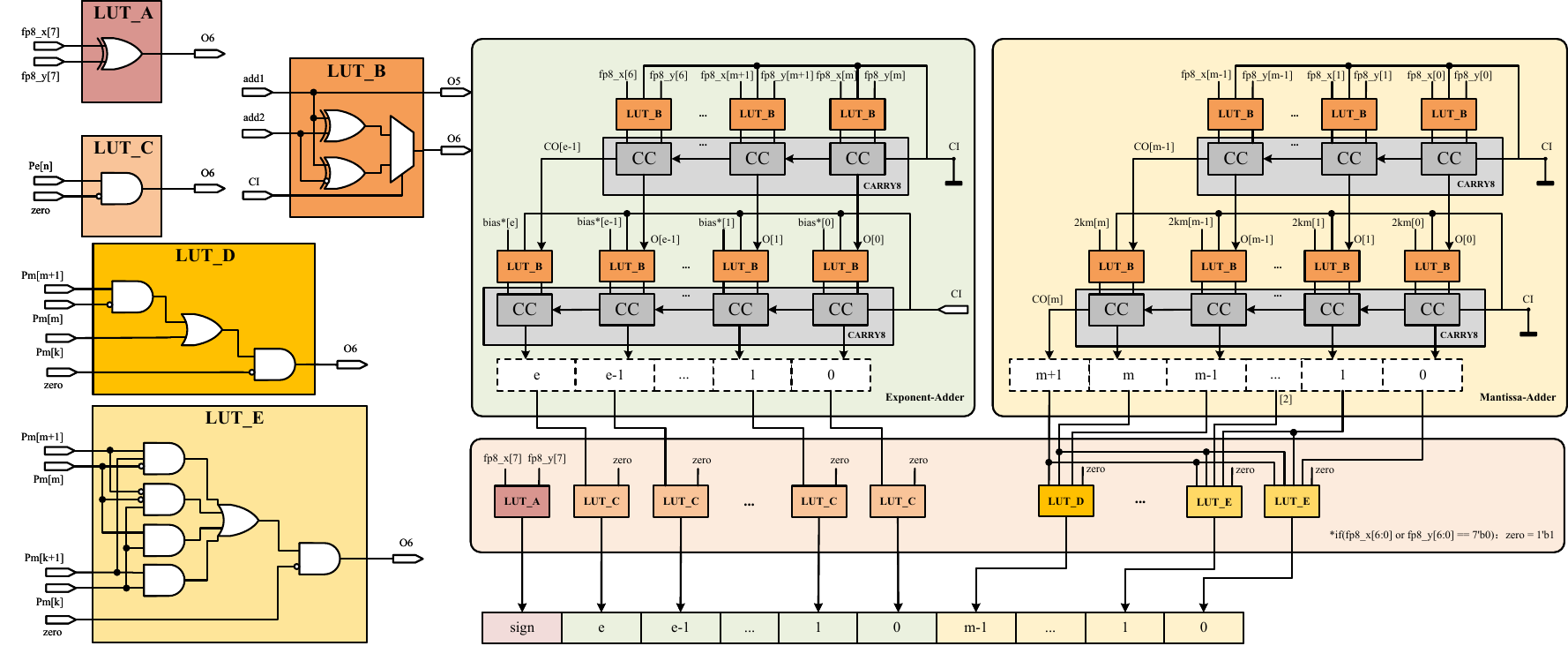}
\caption{Hardware fine-grained architecture for \textit{L-Mul}.}
\label{fig_hardware} 
\end{figure*}

\section{The Hardware Implementation for \textit{L-Mul}}
\subsection{\textit{L-Mul} for FP8}
According the introduction in Section~\ref{sec_bg_fp8}, the FP8 multiplication process can be represented as:

\begin{equation} 	
\begin{split}
Mul(x,y) &= (1+m_{x})\cdot 2^{E_{x}}\times (1+m_{y})\cdot 2^{E_{y}} \\
&=(1+m_{x}+m_{y}+m_{x}\cdot m_{y})\cdot 2^{E_{x}+E_{y}},
\end{split} 
\end{equation}
the sign bit could be omitted, as it can be handled through an $XOR$ operation.
Observing the above equation, it is evident that for hardware circuit design, only $m_{x}\cdot m_{y}$ involves a multiplication operation. The remaining operations can be implemented using addition or linear operations such as shifting. To alleviate the potential bottleneck caused by mantissa multiplication, Luo et al.~\cite{luo2024addition} propose the \textit{L-Mul} algorithm, designed to approximate the FP8 multiplication process:

\begin{equation}
\label{eq_lmul}
\begin{split}
& \text{L-Mul}(x,y) = (1+m_{x}+m_{y}+2^{-l(m)})\times 2^{E_{x}+E_{y}}, \\
& l(m)=\left\{\begin{matrix}   m& \text{if}~m\le3,  \\   3&\text{if}~m=4, \\   4&\text{if}~m\ge 4. \end{matrix}\right.
\end{split} 
\end{equation}
Where, $m$ represents the bit-width of the mantissa. Using this piecewise function approximation, the original multiplication operation can be transformed into shift and addition operations. In the following subsection, we introduce the corresponding fine-grained FPGA-based hardware design.

\subsection{Hardware Design}
 The combination of the \textit{L-Mul} algorithm (Eq.~\ref{eq_lmul}) and the FP8 format conversion relationships (Eq.~\ref{eq_fp8} and Eq.~\ref{eq_bias}) provides the bit-level representation of the \textit{L-Mul} algorithm in binary operations:
\begin{equation}
\label{eq_lmul_bin}
\scalebox{0.9}{$
\begin{aligned}
\text{L-Mul}_{\text{BIN}}(x,y) =& \left(1 + \frac{x[m-1:0] + y[m-1:0] + 2^{l(m)}}{2^{m}} \right) \\
& \times 2^{x[6:m] + y[6:m] - \text{bias}_x - \text{bias}_y}.
\end{aligned}
$}
\end{equation}
To achieve this, we design five LUT configurations combined with carry chains to implement the \textit{L-Mul} FP8 approximate multiplier, as shown in Fig.~\ref{fig_hardware}. The design primarily consists of three parts: the Exponent-Adder, the Mantissa-Adder, and the Post-Processing unit.

The Exponent-Adder and Mantissa-Adder are implemented using $LUT\_B$ and CARRY8. The Exponent-Adder includes an  $m$-bit adder and an  $m$+1-bit adder, while the Mantissa-Adder includes an $e$-bit adder and an $e$+1-bit adder.
$LUT\_B$ primarily functions as a half-adder. The output $O5$ corresponds to the sum ($S$). When the LSB of carry-in ($CI$) is 0, the operation is $O_5=add1\oplus add2$. Otherwise the operation is $O_5=add1\oplus (\sim add2)$. The output ($O_6$) corresponds to the $C$ in a half-adder. In other words, $O_6=add1\cdot add2$. CARRY8 is used to implement addition operations. Each CARRY8 unit contains eight basic units ($CC$), and each $CC$ can combine with $LUT\_B$ to function as a full adder. The $CI$ represents the carry input from the previous stage. When the $CC$ unit is the LSB, $CI = 0$ indicates addition, while $CI = 1$ indicates subtraction. The $O$ corresponds to the sum ($S$) in the full adder, calculated as: $O=(add1\oplus add2)\oplus CI$. In summary, a total of N $LUT\_B$ and $CC$ units can be combined to form an N-bit adder.

\begin{table}[tp!]
\centering
\caption{The representation of the mantissa for different carry}
\label{tab_carry}
\begin{tabular}{|c|c|}
\hline
[m+1,m] & Mantissa \\ \hline
2'b00 & $1.x_m$  \\ \hline
2'b01 & $10.x_m$  \\ \hline
2'b10 & $11.x_m$  \\ \hline
2'b11 & $100.x_m$
\\ \hline
\end{tabular}
\end{table}
The Post-Processing Unit is primarily responsible for handling the sign bit and managing the carry of the mantissa. The $LUT\_A$ is used to determine the sign bit of the product. Specifically, it processes the bit of the inputs $x[7]$ and $y[7]$. There might be a carry occur, requiring the carry value from the mantissa to be added to the exponent. For the mantissa, we follow the carry principles of typical FP multipliers, representing the mantissa in the form of $1.x_m$. The corresponding carry handling is shown in Table~\ref{tab_carry}. When the carry value is $2'b00$, the final product's mantissa is $P_m[m-1:0]$, and no carry is added to the exponent.
 When the carry value is $2'b01$, the mantissa is represented as $10.x_m$, requiring the decimal point to shift left by one position, i.e., the exponent is incremented by 1. In this case, the mantissa is $P_m[m-1:0]$.
 Similarly, when the carry value is $2'b10$, the exponent is incremented by 1, and the mantissa becomes ${1'b1, P_m[m-1:1]}$.
 For a carry value of $2'b11$, the exponent is incremented by 2, and the mantissa is $P_m[m-1:0]$.
 Since the product of 0 and any number is 0, the final product's mantissa and exponent can be expressed using the following formulas
\begin{equation}
\label{eq_pm}
\scalebox{0.83}{$
P'_m[m-1:0] = 
\begin{cases} 
    0, & \text{zero} = 1 \\ 
    \{1'b1, P_m[m-1:1]\}, & P_m[m+1:m] = 2'b10 \\ 
    P_m[m-1:0], & \text{others}
\end{cases}
$}
\end{equation}

\begin{equation}
\label{eq_pe}
P'_e[e:0] = 
\begin{cases} 
    0, & \text{zero} = 1 \\ 
    P_e, & P_m[m+1:m] == 2'b00 \\ 
    P_e + 2, & P_m[m+1:m] == 2'b11 \\ 
    P_e + 1, & \text{others}.
\end{cases}
\end{equation}
To reduce the usage of adders, we combine the bias with various carry scenarios from Table~\ref{tab_carry} and treat it as a constant, $bias*$. The corresponding values are shown in Table~\ref{tab_bias}. $LUT\_C$, $LUT\_D$, and $LUT\_E$ are used to implement Equation~\ref{eq_pm}, ~\ref{eq_pe}, and the remaining corresponding operations. These operations compute the final product's exponent bits, the highest mantissa bit, and the remaining mantissa bits excluding the highest bit, respectively.

\begin{table}[tp!]
\centering
\caption{The $bias*$ values for different types of FP8 formats correspond to specific configurations}
\label{tab_bias}
\begin{tabular}{|c|c|c|c|c|c|}
\hline
\textbf{FP8 Type}    & \textbf{[m+1,m]} & \textbf{bias} & \textbf{FP8 Type}    & \textbf{[m+1,m]} & \textbf{bias} \\ \hline
\multirow{3}{*}{\textbf{E6M1}} & 2'b00 & -31 & \multirow{3}{*}{\textbf{E5M2}} & 2'b00 & -15 \\ \cline{2-3} \cline{5-6}
& 2'b11 & -29 &  & 2'b11 & -13 \\ \cline{2-3} \cline{5-6}
& others & -30 &  & others & -14 \\ \hline

\multirow{3}{*}{\textbf{E4M3}} & 2'b00 & -7 & \multirow{3}{*}{\textbf{E3M4}} & 2'b00 & -3 \\ \cline{2-3} \cline{5-6}
& 2'b11 & -5 &  & 2'b11 & -1 \\ \cline{2-3} \cline{5-6}
& others & -6 &  & others & -2 \\ \hline

\multirow{3}{*}{\textbf{E2M5}} & 2'b00 & -1 & \multirow{3}{*}{\textbf{E1M6}} & 2'b00 & 0 \\ \cline{2-3} \cline{5-6}
& 2'b11 & 1 &  & 2'b11 & 2 \\ \cline{2-3} \cline{5-6}
& others & 0 &  & others & 1 \\ \hline

\end{tabular}
\end{table}

To enhance the performance of the FP8 approximate multiplier, we implemented the hardware design using LUTs and carry chain primitives. Furthermore, to shorten the connection paths between LUTs and carry chains, we applied strict physical placement constraints. Specifically, as described in Section~\ref{sec_FPGA}, each carry chain can connect directly to four LUTs. Therefore, we constrained the physical placement at the CLB level, ensuring that the LUTs are connected to the carry chain within the same CLB. The input FFs
are placed in adjacent CLBs to guarantee the shortest possible data paths.

\section{Results and Discussion}
\subsection{Experimental setup}
 We first evaluate the error of \textit{L-Mul} using five metrics: Error Probability (EP), Mean Absolute Error (MAE), Mean Relative Error (MRE), Mean Squared Error (MSE), and Normalized Error Distance (NED). For unsigned arithmetic, these metrics are defined as follows: 
 \begin{equation}
EP = \frac{1}{2^N} \sum_{i=0}^{2^N-1}ED_i \neq 0,
\end{equation}

\begin{equation}
MAE = \frac{1}{2^N} \sum_{i=0}^{2^N-1} ED_i,
\end{equation}

\begin{equation} 
MRE = \frac{1}{2^N} \sum_{i=0}^{2^N-1} \frac{ED_i}{Exact_i},
\end{equation}

\begin{equation}
MSE = \frac{1}{2^N} \sum_{i=0}^{2^N-1} (ED_i)^2,
\end{equation}

\begin{equation}
NED = \frac{1}{2^N} \sum_{i=0}^{2^N-1} \frac{ED_i}{\max(ED)}.
\end{equation}

 Then, We use Verilog for the \textit{L-Mul} FP8 approximate multiplier design and use AMD Vivado 2022.2 for logic synthesis and placement constraints. The design is deployed and validated on the ZCU104 Evaluation Kit of UltraScale+ FPGA. We perform multiple synthesis iterations, applying different critical path constraints in each iteration to implement each design multiple times. This approach ensures accurate measurements of area and maximum operating frequency. The Vivado simulator and power analysis tools are used to calculate dynamic power consumption. As this is the first FPGA-based FP8 approximate multiplier design, we ensure fairness by selecting previous FPGA-based INT8 approximate multipliers for comparisons of resource consumption, power consumption, and critical path delay. Finally, we deploy the \textit{L-Mul} FP8 approximate multiplier on 2 typical DNN accelerators to validate its superiority in terms of energy efficiency.

\begin{table}[tp!]
\centering
\caption{Error evaluation of \textit{L-Mul} across different FP8 Formats }
\label{tab_error}
\begin{tabular}{|c|c|c|c|c|c|}
\hline
Data type & EP & MAE & MRE & MSE & NED \\ \hline
E6M1 & 1 & 2.1$\times 10^{15}$ & 0.319 & 2$\times 10^{33}$ & 0.001 \\ \hline
E5M2  & 0.938 & 8.58$\times 10^{5}$ & 0.111 & 9.12$\times 10^{13}$ & 0.002 \\ \hline
E4M3  & 0.968 & 141 & 0.068 & 7.56$\times 10^{5}$ & 0.005 \\ \hline
E3M4  & 0.992 & 3.04 & 0.069 & 90.7 & 0.019 \\ \hline
E2M5  & 0.997 & 0.991 & 0.072 & 3.23 & 0.076 \\ \hline
E1M6  & 0.999 & 0.765 & 0.073 & 1.18 & 0.218 \\ \hline
\end{tabular}
\end{table}

\subsection{Error Evaluation}
\begin{figure}[!tp] 
\centering   \includegraphics[width=\linewidth]{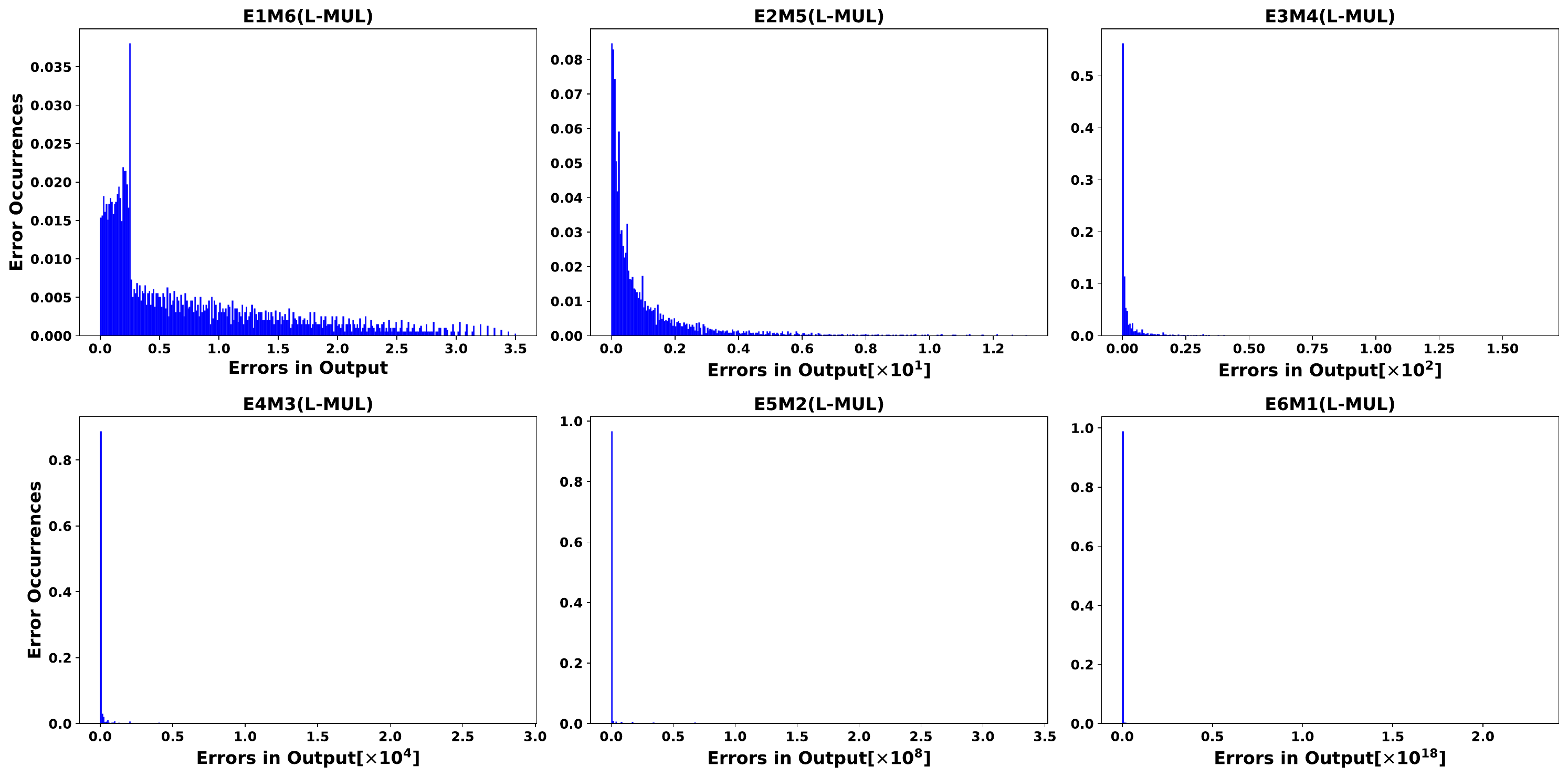}  
\caption{The normalized number of unique error occurrences under different FP8 formats.}   \label{fig_error}   
\end{figure}

Table~\ref{tab_error} presents the error metrics of \textit{L-Mul} in different formats of FP8. It is important to note that when the exponent is allocated a larger bit-width, the range of representable numbers increases, which can lead to significantly larger MAE and MSE values due to the greater magnitude of errors. Moreover, to provide a more intuitive representation of the normalized number of unique error occurrences for the proposed multipliers, we visualize the data in Fig~\ref{fig_error}.

\begin{table}[tp!]
\centering
\caption{Hardware implementation results of \textit{L-Mul} for different FP8 formats}
\label{tb_resource}
\begin{tabular}{|c|c|c|c|c|}
\hline
Data type & LUT & FF & CARRY8/4 &  WNS (ns) \\ \hline
E6M1 & 22 & 25 & 4 & 1.65 \\ \hline
E5M2  & 21 & 25 & 4 & 1.64 \\ \hline
E4M3  & 22 & 25 & 4 & 1.62 \\ \hline
E3M4  & 22 & 25 & 4 & 1.64 \\ \hline
E2M5  & 23 & 25 & 4 & 1.71\\ \hline
E1M6  & 22 & 25 & 4 & 1.76 \\ \hline
\end{tabular}
\end{table}

\subsection{Hardware Implementation and Evaluation} 

\begin{figure}[!tp] 
\centering   \includegraphics[width=\linewidth]{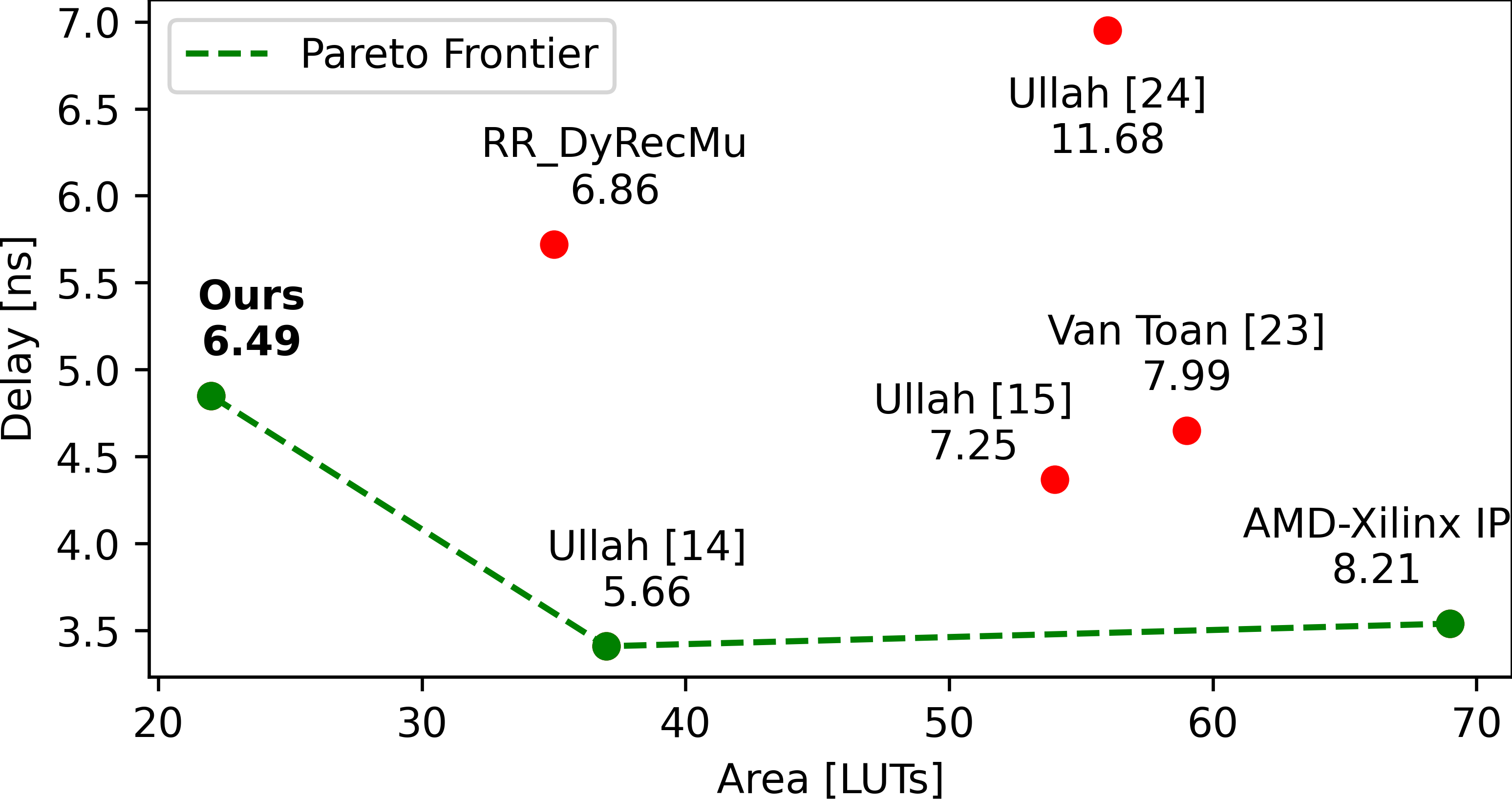}  
\caption{The Pareto analysis under area and latency.}   \label{fig_pareto}   
\end{figure}

We use Verilog to implement the \textit{L-Mul} hardware design for different FP8 formats, as described in Section 3. The corresponding resource consumption and latency for each format is collected from AMD Vivado 2022.2 and shown in Table~\ref{tb_resource}.
It can be observed that our design consumes an average of fewer than 23 LUTs. To further highlight the advantages of our design in terms of resource utilization and power consumption, we compare it with previous FPGA-based approximate multipliers and AMD's IP core. We use the deployment results under the E4M3 format, which is the most commonly used FP8 format. Although the multiplication rules for FP8 and INT8 differ, we consider this comparison valuable due to their identical data bit-width. Table~\ref{tb_compare} presents the comparison results. For each item, it can be observed that our design exhibits the lowest resource consumption. Additionally, compared to the FP8-compatible RR\_DyRecMul~\cite{vakili2024dyrecmul}, our design achieves a lower delay. The fastest frequency, reported in~\cite{van2020fpga}, benefits from the simplicity of unsigned INT8 operations. It is important to note that references~\cite{ullah2018area},~\cite{ullah2020area}, and~\cite{ullah2021high} are implemented on AMD-Xilinx 7-series FPGAs, which introduces some power consumption differences. Additionally, the power data from~\cite{van2020fpga} is recorded at 100 MHz, which is significantly lower than the results from other designs running at their maximum frequencies. To better demonstrate the power-efficiency advantages of our design, we performed a Pareto analysis to visually illustrate the differences between our design and others, as shown in Fig.~\ref{fig_pareto}. It shows that our design lies on the Pareto frontier. Additionally, due to its outstanding power-efficiency, our design achieves the second smallest Power-Delay Product (PDP) among the compared implementations.

\begin{table}[tp!]
\centering
\caption{Implementation Results of Different 8-bit Multipliers}
\label{tb_compare}
\resizebox{\linewidth}{!}{
\begin{tabular}{|c|c|c|c|c|}
\hline
Designs                                                                     & LUTs & Max Frq (MHz) & Delay (ns) & Power (mW) \\ \hline
\begin{tabular}[c]{@{}c@{}}Ours\\ (FP8 E4M3)\end{tabular}  & \textbf{22}     &  617       & 4.85             &   1.34         \\ \hline
\begin{tabular}[c]{@{}c@{}}Ullah~\cite{ullah2018area}\\ (INT8 Unsigned)\end{tabular}             & 56   & /       & 6.95         & 1.68      \\ \hline
\begin{tabular}[c]{@{}c@{}}Van Toan~\cite{van2020fpga}\\ (INT8 Unsigned)\end{tabular}          & 59   & \textbf{759}     & 4.65         & \textbf{0.432}      \\ \hline
\begin{tabular}[c]{@{}c@{}}Ullah~\cite{ullah2020area}\\ (INT8 Unsigned)\end{tabular}             & 37   & /       & \textbf{3.41}         & /          \\ \hline
\begin{tabular}[c]{@{}c@{}}Ullah~\cite{ullah2021high}\\ (INT8 Signed)\end{tabular}             & 54   & /       & 4.37         & 1.66          \\ \hline
\begin{tabular}[c]{@{}c@{}}RR\_DyRecMul~\cite{vakili2024dyrecmul}\\ (FP8 to INT8 Signed)\end{tabular} & 35   & 699     & 5.72         & /          \\ \hline
\begin{tabular}[c]{@{}c@{}}AMD-Xilinx\\ (Exact)\end{tabular}                & 69   & 730     & 3.54         & 2.32      \\ \hline
\end{tabular}}
\end{table}

\subsection{Deployment and Evaluation on Computation of DNN Models}

To further validate the performance and power efficiency of our design, we integrate it into both a CNN accelerator and a GCN accelerator. We evaluate inference accuracy on three representative CNN datasets (MNIST, CIFAR-10, and ImageNetV2) and three typical GCN datasets (CORA, CiteSeer, and Pubmed). Table~\ref{tab_NN} shows the average accuracy loss for the corresponding models under different data formats. FP8, with its superior dynamic range, incurs the least accuracy loss. Although \textit{L-Mul} exhibits the highest accuracy loss, it eliminates multiplication operations, achieving significantly better hardware deployment efficiency. We design CNN inference accelerators based on INT8 and \textit{L-Mul} (FP8, E4M3) using the quantization parameters shown in Table~\ref{tab_NN}, with Faster R-CNN as the backbone network. The resulting resource consumption and power usage are shown in Table~\ref{tab_NNinf}. Thanks to the approximate multiplier design, we achieve a DSP-free implementation and reduce power consumption by 14.59\% at the same operating frequency (250 MHz). Similarly, we implement a GCN inference accelerator based on \textit{L-Mul} (FP8, E4M3) using LW-GCN (INT8)~\cite{tao2022lw}. This design also achieves a DSP-free implementation at the same operating frequency and reduces overall power consumption.

\begin{table}[tp!]
\centering
\caption{Evaluation of Accuracy Loss for Different Data Formats Across Various DNN Models}
\label{tab_NN}
\begin{tabular}{|c|c|c|c|c|}
\hline
Models & FP32 & FP8 (E4M3) & INT8 & \textit{L-Mul} (E4M3) \\ \hline
CNN & 0 & -0.04\% & -0.10\% & -0.96\% \\ \hline
GCN & 0 & -1.96\% & -2.77\% & -3.01\% 
\\ \hline
\end{tabular}
\end{table}

\begin{table}[tp!]
\centering
\caption{Comparison of Hardware Deployment for NN Model Inference Using Different Multipliers}
\label{tab_NNinf}
\begin{tabular}{|c|c|c|c|c|}
\hline
Multiplier                                                                  & Model & LUT     & DSP   & Power (W) \\ \hline
\multirow{2}{*}{\begin{tabular}[c]{@{}c@{}}INT8\\ (Exact)\end{tabular}}     & CNN   & 117,067 & 1,156 & 9.46      \\ \cline{2-5} 
                                                                            & GCN   & 161,529 & 512   & 8.61      \\ \hline
\multirow{2}{*}{\begin{tabular}[c]{@{}c@{}}Ours\\ (FP8, E4M3)\end{tabular}} & CNN   & 143,702 & \textbf{0}     & \textbf{8.08}      \\ \cline{2-5} 
                                                                            & GCN   & 173,387 & \textbf{0}     & \textbf{8.23}      \\ \hline
\end{tabular}
\end{table}

\section{Conclusions}
 This paper presents a power-efficient hardware deployment for the \textit{L-Mul} algorithm. By analyzing the CLB structure of AMD FPGAs, we achieve fine-grained optimization through primitive-based design. We demonstrate and analyze the deployment results for FP8, showing that our design achieves the lowest LUT consumption and power usage compared to other 8-bit designs. Furthermore, we deploy our design in the inference phase of typical NNs, validating its effectiveness and power efficiency. In the future, we plan to integrate this design into LLMs and diffusion model to further demonstrate its advantages.


\begin{thebibliography}{10}
\providecommand{\url}[1]{#1}
\csname url@samestyle\endcsname
\providecommand{\newblock}{\relax}
\providecommand{\bibinfo}[2]{#2}
\providecommand{\BIBentrySTDinterwordspacing}{\spaceskip=0pt\relax}
\providecommand{\BIBentryALTinterwordstretchfactor}{4}
\providecommand{\BIBentryALTinterwordspacing}{\spaceskip=\fontdimen2\font plus
\BIBentryALTinterwordstretchfactor\fontdimen3\font minus \fontdimen4\font\relax}
\providecommand{\BIBforeignlanguage}[2]{{%
\expandafter\ifx\csname l@#1\endcsname\relax
\typeout{** WARNING: IEEEtran.bst: No hyphenation pattern has been}%
\typeout{** loaded for the language `#1'. Using the pattern for}%
\typeout{** the default language instead.}%
\else
\language=\csname l@#1\endcsname
\fi
#2}}
\providecommand{\BIBdecl}{\relax}
\BIBdecl

\bibitem{dampfhoffer2023backpropagation}
M.~Dampfhoffer, T.~Mesquida, A.~Valentian, and L.~Anghel, ``Backpropagation-based learning techniques for deep spiking neural networks: A survey,'' \emph{IEEE Transactions on Neural Networks and Learning Systems}, vol.~35, no.~9, pp. 11\,906--11\,921, 2024.

\bibitem{luo2024addition}
H.~Luo and W.~Sun, ``Addition is all you need for energy-efficient language models,'' \emph{arXiv preprint arXiv:2410.00907}, 2024.

\bibitem{han2020extremely}
Q.~Han, Y.~Hu, F.~Yu, H.~Yang, B.~Liu, P.~Hu, R.~Gong, Y.~Wang, R.~Wang, Z.~Luan, and D.~Qian, ``Extremely low-bit convolution optimization for quantized neural network on modern computer architectures,'' in \emph{Proceedings of the 49th International Conference on Parallel Processing}, ser. ICPP '20.\hskip 1em plus 0.5em minus 0.4em\relax New York, NY, ACM, 2020.

\bibitem{yao2022zeroquant}
Z.~Yao, R.~Yazdani~Aminabadi, M.~Zhang, X.~Wu, C.~Li, and Y.~He, ``Zeroquant: Efficient and affordable post-training quantization for large-scale transformers,'' in \emph{Advances in Neural Information Processing Systems}, vol.~35.\hskip 1em plus 0.5em minus 0.4em\relax Curran Associates, Inc., 2022, pp. 27\,168--27\,183.

\bibitem{shen2024efficient}
H.~Shen, N.~Mellempudi, X.~He, Q.~Gao, C.~Wang, and M.~Wang, ``Efficient post-training quantization with fp8 formats,'' in \emph{Proceedings of Machine Learning and Systems}, P.~Gibbons, G.~Pekhimenko, and C.~D. Sa, Eds., vol.~6, 2024, pp. 483--498.

\bibitem{lutz2024fused}
D.~R. Lutz, A.~Saini, M.~Kroes, T.~Elmer, and H.~Valsaraju, ``Fused fp8 4-way dot product with scaling and fp32 accumulation,'' in \emph{2024 IEEE 31st Symposium on Computer Arithmetic (ARITH)}.\hskip 1em plus 0.5em minus 0.4em\relax IEEE, 2024, pp. 40--47.

\bibitem{lee20217}
S.~K. Lee, A.~Agrawal, J.~Silberman, M.~Ziegler, M.~Kang, S.~Venkataramani, N.~Cao, B.~Fleischer, M.~Guillorn, M.~Cohen \emph{et~al.}, ``A 7-nm four-core mixed-precision ai chip with 26.2-tflops hybrid-fp8 training, 104.9-tops int4 inference, and workload-aware throttling,'' \emph{IEEE Journal of Solid-State Circuits}, vol.~57, no.~1, pp. 182--197, 2021.

\bibitem{venkataramanaiah202328}
S.~K. Venkataramanaiah, J.~Meng, H.-S. Suh, I.~Yeo, J.~Saikia, S.~K. Cherupally, Y.~Zhang, Z.~Zhang, and J.-S. Seo, ``A 28-nm 8-bit floating-point tensor core-based programmable cnn training processor with dynamic structured sparsity,'' \emph{IEEE Journal of Solid-State Circuits}, vol.~58, no.~7, pp. 1885--1897, 2023.

\bibitem{elster2022nvidia}
A.~C. Elster and T.~A. Haugdahl, ``Nvidia hopper gpu and grace cpu highlights,'' \emph{Computing in Science \& Engineering}, vol.~24, no.~2, pp. 95--100, 2022.

\bibitem{ultra}
\BIBentryALTinterwordspacing
AMD Xilinx. Ultrascale architecture configurable logic block user guide. Accessed: 2024-11-11. [Online]. Available: \url{https://docs.amd.com/v/u/en-US/ug574-ultrascale-clb}
\BIBentrySTDinterwordspacing

\bibitem{micikevicius2022fp8}
P.~Micikevicius, D.~Stosic, N.~Burgess, M.~Cornea, P.~Dubey, R.~Grisenthwaite, S.~Ha, A.~Heinecke, P.~Judd, J.~Kamalu \emph{et~al.}, ``Fp8 formats for deep learning,'' \emph{arXiv preprint arXiv:2209.05433}, 2022.

\bibitem{van2023fp8}
M.~van Baalen, A.~Kuzmin, S.~S. Nair, Y.~Ren, E.~Mahurin, C.~Patel, S.~Subramanian, S.~Lee, M.~Nagel, J.~Soriaga \emph{et~al.}, ``Fp8 versus int8 for efficient deep learning inference,'' \emph{arXiv preprint arXiv:2303.17951}, 2023.

\bibitem{ansari2020improved}
M.~S. Ansari, B.~F. Cockburn, and J.~Han, ``An improved logarithmic multiplier for energy-efficient neural computing,'' \emph{IEEE Transactions on Computers}, vol.~70, no.~4, pp. 614--625, 2020.

\bibitem{zheng2022heam}
S.~Zheng, Z.~Li, Y.~Lu, J.~Gao, J.~Zhang, and L.~Wang, ``Heam: High-efficiency approximate multiplier optimization for deep neural networks,'' in \emph{2022 IEEE International Symposium on Circuits and Systems (ISCAS)}.\hskip 1em plus 0.5em minus 0.4em\relax IEEE, 2022, pp. 3359--3363.

\bibitem{chen2020optimally}
C.~Chen, S.~Yang, W.~Qian, M.~Imani, X.~Yin, and C.~Zhuo, ``Optimally approximated and unbiased floating-point multiplier with runtime configurability,'' in \emph{Proceedings of the 39th international conference on computer-aided design {(ICCAD)}}, 2020, pp. 1--9.

\bibitem{ullah2018smapproxlib}
S.~Ullah, S.~S. Murthy, and A.~Kumar, ``Smapproxlib: Library of fpga-based approximate multipliers,'' in \emph{Proceedings of the 55th Annual Design Automation Conference {(DAC)}}.\hskip 1em plus 0.5em minus 0.4em\relax New York, NY, ACM, 2018, pp. 1--6.

\bibitem{ullah2020area}
S.~Ullah, H.~Schmidl, S.~S. Sahoo, S.~Rehman, and A.~Kumar, ``Area-optimized accurate and approximate softcore signed multiplier architectures,'' \emph{IEEE Transactions on Computers}, vol.~70, no.~3, pp. 384--392, 2020.

\bibitem{ullah2021high}
S.~Ullah, S.~Rehman, M.~Shafique, and A.~Kumar, ``High-performance accurate and approximate multipliers for fpga-based hardware accelerators,'' \emph{IEEE Transactions on Computer-Aided Design of Integrated Circuits and Systems}, vol.~41, no.~2, pp. 211--224, 2021.

\bibitem{liu2024bitsys}
Y.~Liu, S.~Ullah, and A.~Kumar, ``Bitsys: Bitwise systolic array architecture for multi-precision quantized hardware accelerators,'' in \emph{2024 IEEE 32nd Annual International Symposium on Field-Programmable Custom Computing Machines (FCCM)}.\hskip 1em plus 0.5em minus 0.4em\relax IEEE, 2024, pp. 220--220.

\bibitem{ullah2024axospike}
S.~Ullah, S.~S. Sahoo, and A.~Kumar, ``Axospike: Spiking neural networks-driven approximate operator design,'' \emph{IEEE Transactions on Computer-Aided Design of Integrated Circuits and Systems}, vol.~43, no.~11, pp. 3324--3335, 2024.

\bibitem{vakili2024dyrecmul}
S.~Vakili, M.~Vaziri, A.~Zarei, and J.~P. Langlois, ``Dyrecmul: Fast and low-cost approximate multiplier for fpgas using dynamic reconfiguration,'' \emph{ACM Transactions on Reconfigurable Technology and Systems}, 2024.

\bibitem{leon2021improving}
V.~Leon, T.~Paparouni, E.~Petrongonas, D.~Soudris, and K.~Pekmestzi, ``Improving power of dsp and cnn hardware accelerators using approximate floating-point multipliers,'' \emph{ACM Transactions on Embedded Computing Systems (TECS)}, vol.~20, no.~5, pp. 1--21, 2021.

\bibitem{ullah2018area}
S.~Ullah, S.~Rehman, B.~S. Prabakaran, F.~Kriebel, M.~A. Hanif, M.~Shafique, and A.~Kumar, ``Area-optimized low-latency approximate multipliers for fpga-based hardware accelerators,'' in \emph{Proceedings of the 55th Annual Design Automation Conference}, ser. DAC '18, 2018.

\bibitem{van2020fpga}
N.~Van~Toan and J.-G. Lee, ``Fpga-based multi-level approximate multipliers for high-performance error-resilient applications,'' \emph{IEEE Access}, vol.~8, pp. 25\,481--25\,497, 2020.

\bibitem{tao2022lw}
Z.~Tao, C.~Wu, Y.~Liang, K.~Wang, and L.~He, ``Lw-gcn: A lightweight fpga-based graph convolutional network accelerator,'' \emph{ACM Transactions on Reconfigurable Technology and Systems}, vol.~16, no.~1, pp. 1--19, 2022.

\end{thebibliography}
\end{document}